\documentclass[a4paper,fleqn]{cas-dc}

\usepackage[numbers]{natbib}

\def\tsc#1{\csdef{#1}{\textsc{\lowercase{#1}}\xspace}}
\tsc{WGM}
\tsc{QE}
\tsc{EP}
\tsc{PMS}
\tsc{BEC}
\tsc{DE}

\begin{document}
\let\WriteBookmarks\relax
\def\floatpagepagefraction{1}
\def\textpagefraction{.001}
\shorttitle{The mass of the neutron star in 4U 1820–30 revisited}
\shortauthors{M G B de Avellar et~al.}

\title [mode = title]{The mass of the neutron star in 4U 1820–30 revisited}                      
\tnotemark[1]

\tnotetext[1]{This document is the results of the research
   project funded by the São Paulo Research Foundation (FAPESP) through the grants \#2025/05690-2, \#2023/08649-8. and \#2024/16892-2; Alexander von Humboldt Foundation; Brazilian Ministry of Science, Technology and Innovation (MCTI) and the Brazilian Space Agency (AEB), under the PO 20VB.0009; the CNPq Federal Agency. }


\author[1]{Marcio G B de Avellar}[type=editor,
                        orcid=0000-0003-1771-1055]
\cormark[1]
\ead{marcio.avellar@inpe.br}

\credit{Writing – original draft, Writing – review \& editing, Visualization, Software, Formal analysis, Methodology}

\affiliation[1]{organization={Instituto Nacional de Pesquisas Espaciais (INPE)},
                addressline={Av. dos Astronautas 1758}, 
                city={São José dos Campos},
                postcode={12227-010}, 
                state={São Paulo},
                country={Brazil}}

\author[2]{Jorge E Horvath}[type=editor,
                        orcid=0000-0003-4089-3440]
\cormark[1]
\ead{foton@iag.usp.br}

\credit{Conceptualization, Writing – review \& editing, Methodology}

\affiliation[2]{organization={Instituto de Astronomia, Geofísica e Ciências Atmosféricas, Universidade de São Paulo (USP)},
                addressline={R. do Matão, 1226, Cidade Universitária}, 
                city={São Paulo},
                postcode={05508-090}, 
                state={São Paulo},
                country={Brazil}}

\author[2]{Nathalia D Pires}[type=editor,
                        orcid=0009-0000-7480-5395]
\credit{Writing - Citation curation, comments}

\author[3]{Lívia S Rocha}[type=editor,
                        orcid=0000-0003-4543-0912]
\credit{Writing – review \& editing}
\affiliation[3]{organization={Instituto de Física, Universidade de São Paulo (USP)},
                addressline={R. do Matão, 1371, Cidade Universitária}, 
                city={São Paulo},
                postcode={05508-090}, 
                state={São Paulo},
                country={Brazil}}

\author[4]{Lucas M de Sá}[type=editor,
                        orcid=0000-0003-3109-9042]
\credit{Writing - Citation curation, comments}
\affiliation[4]{organization={Universit\"at Heidelberg, Zentrum f\"ur Astronomie, Institut f\"ur Theoretische Astrophysik},
                addressline={Albert Ueberle Str. 2}, 
                city={Heidelberg},
                postcode={69120}, 
                country={Germany}}

\author[2]{Bianca B Martins}[type=editor,
                        orcid=0009-0002-9944-9889]
\credit{Writing - Citation curation, comments}

\author[2]{Lucas G Barão}[type=editor,
                        orcid=0009-0000-3268-9058]
\credit{Writing - Citation curation, comments}

\author[2]{Davi V Rodrigues}[type=editor,
                        orcid=0009-0008-0984-7642]
\credit{Writing - Citation curation, comments}

\author[2]{Guilherme R C Sampaio}[type=editor,
                        orcid=0009-0008-7350-7546]
\credit{Writing - Citation curation, comments}

\author[2]{Rachel M Takahashi}[type=editor,
                        orcid=0009-0007-7430-8146]
\credit{Writing - Citation curation, comments}

\cortext[cor1]{Corresponding author}

\begin{abstract}
We revisit the mass of the neutron star in the ultracompact binary 4U 1820--30 in light of a recently reported transient absorption feature at about 3.8 keV, interpreted as a gravitationally redshifted, highly ionized iron line and implying \(1+z\simeq1.72\), a very high stellar compactness. We examine whether the mass-radius locus implied by this interpretation can be made compatible with external EoS-informed benchmarks and timing-based estimates. We map the compactness implied by the redshift onto the mass-radius plane, including rotational effects, and compare the resulting region with an EoS-informed 95\% reference contour derived from NICER data of several neutron-star systems. While we do not perform a statistically self-consistent joint mass-radius inference for 4U 1820--30, we present a quantitative conditional consistency test in a common \(M-R\) framework, comparing published inputs under explicitly stated assumptions. We find that, if the maximum neutron-star mass is restricted to low values \((\leq 2.3\,M_{\odot})\), the redshift-implied locus shows at most marginal overlap with the EoS-informed contour, indicating substantial tension. Allowing a higher maximum mass enlarges the parameter space and can restore compatibility with that benchmark. However, such high masses remain in tension with previous touchdown-flux estimates, although they are not necessarily excluded by interpretations based on the highest detected quasi-periodic oscillation frequency. We conclude that the redshift interpretation of the 3.8 keV feature, the touchdown-flux estimates, and the QPO/ISCO interpretation do not naturally select the same mass-radius sector for this source under these assumptions. Reconciling them requires auxiliary assumptions with high leverage on the inferred compactness.
\end{abstract}




\maketitle

\section{Introduction}\label{introduction}

The Low-Mass X-Ray Binary (LMXB) system 4U 1820-30 has been known for many decades and is an extreme system of this class, with an orbital period of $\sim 11 \, \mathrm{min}$ (in fact, among the shortest currently known for this class \citep{StellaL_1987ApJ...315L..49S}). The companion white dwarf is very light ($  M\simeq 0.07 \, M_{\odot}$; \citep{StellaL_1987ApJ...315L..49S, verbunt1987, AndersonS_1997ApJ...482L..69A}), with the neutron star accreting matter and undergoing thermonuclear Type~I X-ray bursts episodically. Burst spectroscopy has revealed absorption lines \citep{FutamotoK_2004ApJ...605..793F,CackettE_2008ApJ...677.1233C},  and the ``touchdown flux'' method has been used to infer both the mass and radius \citep{ozel_2006,suleimanov2011} of the neutron star. A non-negligible fraction of the mass lost by the companion may have contributed to increasing the neutron star mass, given its weak magnetization \citep{DiMarcoA_2025AN....34640126D,BobrikovaA_2025AN....34640128B}. 
As we discuss below, an early source-specific touchdown-flux analysis by \citet{GuverT_2010ApJ...719.1807G} reported a narrow mass estimate for 4U 1820–30, \(M=1.58\pm0.06\,M_\odot\), within the assumptions of the method; later (burst-based) works revised this estimate upward and increased the associated uncertainty \citep{OzelF_2016ApJ...820...28O,OzelF_2016ARA&A..54..401O}.

In Sections \ref{newconstraint} and \ref{section:Results}, we assess the recent NICER-based mass inference from a transient absorption feature at $\sim$3.8~keV. This feature is interpreted as a gravitationally redshifted highly ionized Fe line formed at (or very near) the neutron star surface, yielding $1+z\simeq1.72$, and therefore a very high compactness through the usual expression $1+z=[1-2GM/(c^2R)]^{-1/2}$ \citep{IariaR_2025arXiv251120499I}. If confirmed, this would directly constrain the mass--radius {\it locus}. We show that the compactness implied by this interpretation is difficult to reconcile with EOS-informed benchmarks and timing-based estimates unless one allows assumptions that substantially reshape the admissible \(M\)--\(R\) sector. We shall discuss the results and draw some conclusions in Sections \ref{discussion} and \ref{conclusions}.

\section{Previous determinations of the mass of 4U 1820-30 neutron star}\label{previousDeterminations}

\subsection{The mass of the neutron star in 4U 1820--30 from thermonuclear-burst spectroscopy}

The standard determination of compact object masses in binaries is well-known, and we refer to \citealt{HorvathJ_2017hsn..book.1317H} for a review. In practice, the orbital inclination $i$ is often the dominant source of uncertainty because the inferred masses strongly depend on $\sin i$ through the binary mass function \citep{SaL_2023Galax..11...19D,RochaL_2023Univ...10....3R}. In addition, the mass ratio $q$ may be uncertain and loosely constrained. In the case of the 4U 1820-30 source there is a consensus that the companion white dwarf mass is low, $M_{\rm WD}\simeq 0.055$--$0.07\,M_\odot$ \citep{StellaL_1987ApJ...315L..49S,verbunt1987,AndersonS_1997ApJ...482L..69A}, but the neutron star mass is not well constrained under the simplest assumptions of this construction. In fact, many papers have assumed a neutron star mass of $1.4 M_{\odot}$ in the absence of a robust source-specific measurement.


Besides the mass-function studies, the touchdown-flux framework, as implemented in earlier applications by \citet{ozel_2006} and \citet{OzelF_2009ApJ...693.1775O}, became a widely used approach to infer neutron-star masses and radii from photospheric-radius-expansion thermonuclear X-ray bursts. Specifically for source 4u~1820--30, \citet{GuverT_2010ApJ...719.1807G} obtained a narrow estimate of \(M=1.58\pm0.06\, M_\odot\) under the assumptions of the framework.

The so-called ``touchdown flux'' method \citep{ozel_2006} assumes that the touchdown moment corresponds to the photosphere returning to the neutron star surface \((r_{ph} = R\), where $r_{ph}$ is the radius of the photosphere), and therefore, that the measured flux is the Eddington flux at the surface. Later studies showed that this condition is not observationally guaranteed and can introduce significant biases in the analysis. \cite{Steiner_etal_2010} found extremely low acceptance rates for physically real solutions for neutron star mass and radius in their Monte Carlo (MC) analyses, suggesting inconsistencies with the observables once uncertainties are properly propagated.

Despite this seemingly encouraging result for the source, several studies have raised some concerns about a robust derivation of $M$ and $R$ from burst spectroscopy. The criticisms include systematic uncertainties associated with three main points, namely, the true atmosphere composition and spectral color-correction factors; the deviations from spherical symmetry and isotropic emission; and the potential biases introduced by the burst selection and modeling choices (see, e.g., \cite{Steiner_etal_2010}). The latest determination using the same method \citep{OzelF_2016ApJ...820...28O,OzelF_2016ARA&A..54..401O} yielded the value $1.77 {^{+0.25}_{-0.28}} M_{\odot}$ for the neutron star mass, correcting the original result. It should be mentioned that other works \citep{Kuulkers, suleimanov} obtained alternative values for the neutron star mass, illustrating the difficulty of this extraction.

A more recent work \citep{kim_etal_2021}, using MC analyses and the parameter \(h = 2R / r_{ph}\) from \cite{Steiner_etal_2010}, attempted to determine the touchdown radius as a function of the NS radius, and showed that for 4U 1820–30 the posterior distribution of $h$ does {\it not} concentrate around \(h \approx 2\), as expected for the touchdown case \( R = r_{ph} \) proposed by \cite{ozel_2006}, but instead favors smaller values, implying \( r_{ph} > R \) and often \( r_{ph} > 2R \). This supports the interpretation that the touchdown condition actually occurs above the NS surface. Moreover, although 4U 1820–30 favors a low-hydrogen mass fraction $X$ (the H-He fraction in the plasma given the nature of its donor), the joint behavior of $X$ and $h$ is quite “atypical”, and suggests that simplifications in the assumed composition and accretion physics may break down precisely where the method is expected to be more reliable.

In this context, \citet{kim_etal_2021} have shown that inferred radii can come out systematically larger under alternative (and arguably more realistic) treatments of these systematics. Simulations reported therein support the same trend. If this bias is present, then the mass inference for 4U~1820--30 (and possibly for similar systems) may be affected at a level that is non-negligible for our purposes. We do not reanalyse the burst data here. Instead, we use the corrected burst-based estimates as previous source-specific reference values, with their own systematic limitations, against which the redshift- and QPO-based constructions can be compared.

\subsection{The mass of the 4U 1820-30 neutron star determined using the QPOs of the spectrum}

Quasi-periodic oscillations of the X-ray light curves have been detected almost 50 years ago for the first time \citep{discovery}. Besides the low-frequency range ones, at high values the sources generally show a pair of close peaks with varying degrees of coherence. The separation of these peaks remains more or less constant in some cases, and the attempts to interpret them as signatures of the innermost stable circular orbit (ISCO) led to a long-standing debate \citep{Mariano, Kaaret} about the viability of the interpretation.

For the specific case of the source 4U 1820-30, the suggestion that the frequencies correlate with the mass accretion rate prompted the search of correlation with the count rate, later found to be an unreliable indicator of the accretion rate. However, detailed observations of the source \citep{Kaaret} found that the frequencies do correlate with the energy flux and X-ray spectral shape, saturating at some value of these quantities. Interpreted as evidence that a minimum radius has been achieved for the orbits of the emitting gas, the frequencies are seen as evidence for the ISCO of inspiraling matter. 

Under the simplifying assumption that the highest frequency traces the ISCO, one obtains the commonly used scaling

\begin{equation}
    M_{NS} = 2.2 \,\, {\bigg( {\frac{1000 Hz}{\nu_{QPO}}} \biggr)} \,\, M_{\odot}
\end{equation}

In this picture, the highest frequency is assumed to correspond to the innermost orbit ISCO. For the case of 4U 1820-30 the upper value is $\sim 1065 \, Hz$, which results in a value $M_{NS} > 2 M_{\odot}$. This is actually a  {\it conditional lower limit} to the mass of the neutron star in 4U 1820-30, since it is known that the spacetime is affected by the spin of the central object, identified with a candidate burst-oscillation frequency and measured to be $1.4 \, ms$ (\cite{spinMeasurement}). To account for this rotation effect, at first order a correction of $(1+0.75 j)$ (with $j$ the dimensionless quotient of the angular momentum contribution to the energy for an assumed prograde orbit of the gas) may be assumed, which results in a small positive value that yields $M > 2 M_{\odot}$, conservatively stated.

Given these figures, this estimate is in clear tension with the touchdown-flux determinations and points toward a comparatively massive neutron star. This would be important for the context of the latest report on the source given in the next Section.

For this specific source 4U 1820-30, \cite{1997ApJ...483L.119S} and \cite{Kaaret} detected an upper kilo-hertz QPO at \(1065\pm 7\) Hz. Here we assumed this frequency as the frequency at the ISCO and calculated the set of masses and radii that satisfy this constraint.

The mass value of a rapidly rotating central object can be refined using a treatment beyond the first order correction factor \(1+0.75j\). For this refined calculation, we first used the full expression in the Kerr metric below \citep{2018grav.book.....M}

\begin{equation}
\begin{split}
\nu_{K, ISCO} = \frac{c^{3}}{2\pi6^{3/2}GM}\Bigg(1-\frac{\sqrt{6}}{36}j\Bigg(\frac{1}{1-\frac{2\sqrt{6}}{9}j}\Bigg)^{3/2}\Bigg)\quad\times \\ \Bigg(1-\frac{2\sqrt{6}}{9}j\Bigg)^{-3/2}.
\end{split}
\end{equation}

\noindent where \(j\equiv \frac{cJ}{GM^{2}}=\frac{cI\omega}{GM^{2}}\), \(I\equiv\frac{2}{5}MR^{2}\) is the momentum of inertia of a rigid sphere, and \(\omega = 2\pi f\),  with \(f=716\) Hz the adopted spin frequency of the neutron star in 4U 1820-30 \citep{spinMeasurement}. 

The computation is now quite straightforward: we impose \(\nu_{K, ISCO}=1065\) and for each given mass we find the radius $R$ that satisfies the expression. Conservatively, as a necessary condition for physical admissibility, we impose the Buchdahl bound \citep{Buchdahl}, \(R >\frac{9GM}{4c^{2}}\). For static, spherically symmetric perfect-fluid configurations, this bound is saturated by the uniform-density (incompressible) solution. Moreover, knowing that the exterior Kerr metric is accurate for a black hole solution, we also used the Hartle \& Thorne metric \citep{HT,2017bhns.work...23G,2003gr.qc....12070A} which is more suitable for relativistic stars. In this case:

\begin{equation}
\begin{split}
\nu_{K, ISCO}=\frac{c^{3}}{2\pi6^{3/2}GM} \times 
 \Bigg[1+\frac{11}{6\sqrt{6}}j + \\ \frac{1}{864}\Big(-160583+397710\ln{\frac{3}{2}}\Big)j^{2}
+ \\ \frac{5}{32}\Big(1193-2946\ln{\frac{3}{2}}\Big)q\Bigg]
\end{split}
\end{equation}

\noindent where \(q\) is the dimensionless quadrupole parameter. We adopt the parametrization \(q=-k j^{2}\), with \(k\in[1,8]\). In this notation, \(k=1\) corresponds to the Kerr quadrupole, whereas \(k>1\) mimics the larger quadrupolar deformation expected for realistic neutron stars.

The parameter \textit{k} reflects the response of the star to rotation. Physically, it measures how much the gravitational quadrupole of the star departs from the Kerr case for the same mass $M$ and angular momentum $J$. The Kerr limit is \(k=1\), whereas \(k >1\) represents a star that is both more deformable and more oblate, as expected for realistic neutron stars. Indirectly, it encodes the internal structure of the star and its equation of state. Once these curves are obtained, we plotted the results in Figure \ref{fig:Locus}.

The estimate of \(j\) is used here to assess the scale and direction of spin-related corrections in the \(M\)--\(R\) plane. It is not meant to replace a fully self-consistent EOS-dependent calculation of rapidly rotating neutron-star configurations. Nevertheless, the Kerr and Hartle--Thorne ISCO expressions capture the leading relativistic dependence needed for the present comparison and allow us to test whether rotational corrections can shift the redshift-implied locus by the required amount. The resulting curves should therefore be read as controlled approximate diagnostics, not as precision rotating-star solutions.

The QPO/ISCO curves in Figure \ref{fig:Locus} play a conditional role. If the upper kHz QPO at 1065 Hz is associated with the ISCO, they show where the redshift-based \(M\)--\(R\) locus remains compatible with that timing interpretation. Since the QPO--ISCO identification is model-dependent, and the adopted 716 Hz spin is tied to a candidate burst-oscillation frequency, these curves should not be treated as model-independent exclusions. They are nevertheless useful because they expose the tension between timing-based and redshift-based mass estimates.

\section{A new redshift constraint mapped into the \texorpdfstring{\(M-R\)}~ plane}\label{newconstraint}

As it stands, the issue of the neutron star mass in 4U 1820-30 is still not firmly established and further work is needed to clarify it. Up to now, we have shown the determinations from the ``touchdown flux'' and the 
QPOs analysis, but other methods are potentially available.
Recently, \cite{IariaR_2025arXiv251120499I} reported a transient absorption feature at $\sim3.8$ keV in NICER spectra of the LMXB 4U~1820--30, temporally close to a carbon superburst detected by MAXI. The authors interpreted the feature as a gravitationally redshifted, highly ionized Fe absorption line, inferring $1+z\simeq 1.72$ from a photoionization absorption fit. Within this interpretation, they used the feature to calculate the compactness $M/R$. If the line identification and formation site are correct, this interpretation could in principle provide a strong constraint on the neutron-star equation of state.  

The physics behind the feature led the authors to argue that the superburst-driven phase temporarily thins or removes the Comptonizing corona, exposing the neutron star photosphere. As the wind fades and the photosphere relaxes, the continuum crosses the thin, highly ionized atmosphere, which enables the detection of the broad Fe absorption line (which then weakens as the corona is reestablished). 

Notably, \cite{IariaR_2025arXiv251120499I} acknowledged that the inference depends on the post-superburst atmospheric composition, and that alternative identifications exist in the literature (e.g., Ar XVIII/Ca XX from the inner disk; neutral Ti or H-like Ca). All these would imply more moderate redshifts.  

Here we treated the redshift reported by \cite{IariaR_2025arXiv251120499I} as an \emph{input} to the $M$--$R$ plane under clearly stated auxiliary assumptions. We test the geometric compatibility of the implied locus of the compactness with independent constraints under different $M_{\max}$ assumptions, rather than re-fitting the spectrum.

In the static limit (using the relativistic units $G=c=1$), the gravitational redshift of photons emitted from the radius $R$ is  

\begin{equation}
1+z \;=\; \left(1-\frac{2M}{R}\right)^{-1/2} ~ \Rightarrow ~ \frac{R}{M} \;=\; \frac{2}{1-(1+z)^{-2}}.
\label{eq:RMofz}
\end{equation}

Their best-fit yields $1+z=1.72\pm0.05$ (99.7\% confidence), corresponding to $R/M = 4.46 \pm 0.13\,km/M_{\odot}$. 

As it stands, Equation \ref{eq:RMofz} defines a one-parameter locus in the $M$--$R$ plane: for any assumed $M$, this yields $R=f(z)\,M$, where \emph{f} is a function of the redshift \emph{z}. \cite{IariaR_2025arXiv251120499I} used this locus to illustrate the implied compactness by simple substitutions (e.g., $M=1.4\,M_\odot \Rightarrow R\simeq 6.25$ km; $R=10$ km $\Rightarrow M\simeq 2.24\,M_\odot$). Furthermore, they adopted a spin frequency of 716 Hz \citep{spinMeasurement} for the neutron star, and an orbital inclination of $i=31^\circ$ \citep{AnitraA_2025A&A...697A..83A} to compute rotating-star $M(R)$ solutions for $(1+z)=1.72\pm0.05$ at 716 Hz. The $M(R)$ solutions were fitted with a quadratic relation shown in Equation ~\ref{eq:MofRnum}, which we reproduce as the ``rotation-corrected'' locus. We treat the redshift uncertainty band (blue shading in their Fig.~9 caption) as the dominant propagated uncertainty on this locus.

\begin{equation}
\begin{split}
\frac{M}{M_{\odot}} =
(-1.64 \pm 0.03)\times10^{-3}\Bigg(\frac{R}{1\,km}\Bigg)^{2}
 + \\ (0.240\pm0.007)\Bigg(\frac{R}{1\,km}\Bigg) - (0.0062\pm 0.001).\label{eq:MofRnum}
\end{split}
\end{equation}

We define the ``rotation wedge'' as the region in the $M$--$R$ plane spanned by the rotation-corrected best-fit curve and its redshift-uncertainty band. We computed it using the same $(\nu,i)$ and the same metric/structure approximations as \cite{IariaR_2025arXiv251120499I}. Then, we overlaid this construction against independent constraints to evaluate their compatibility (see Figure \ref{fig:Locus} in section \ref{section:Results}).

\section{Results}
\label{section:Results}

\begin{figure} 
\centering
\includegraphics[width=0.95\columnwidth]{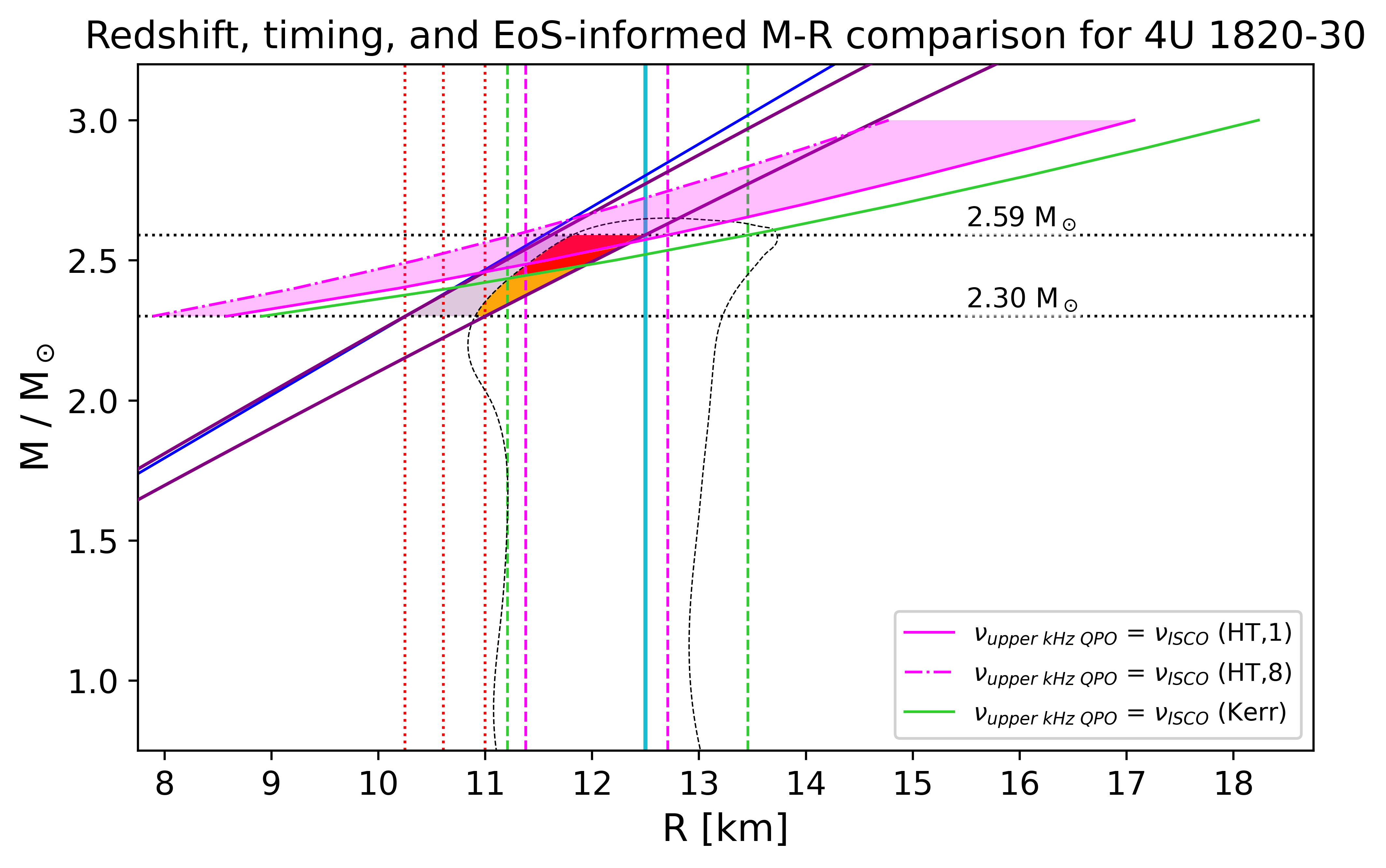}
\caption{Conditional \(M\)--\(R\) consistency map for 4U 1820--30 under the adopted assumptions. The irregularly dotted contour shows the 95\% EOS-informed reference band derived from NICER measurements of neutron-star masses \(M\) (in solar units) and radii \(R\) (in km). This NICER-based contour is included only as an EOS-informed benchmark and should not be interpreted as a direct observational constraint on 4U 1820--30. The horizontal black dotted lines mark two assumed values of the EOS maximum mass, \(M_{\max}=2.30\,M_{\odot}\) \citep{IariaR_2025arXiv251120499I} and \(M_{\max}=2.59\,M_{\odot}\) \citep{RochaL_2023Univ...10....3R}. The upper dark-blue diagonal line corresponds to the best-fit inverse compactness \(R/M = 4.46 \pm 0.13\), inferred from a redshift \(1+z=1.72 \pm 0.05\) (99.7\% confidence) for a non-rotating neutron star. The two purple diagonal lines bracket the redshift uncertainty when approximate rotational corrections for the adopted candidate spin frequency of \(716\,\mathrm{Hz}\) are included. The pink-shaded region marks the corresponding conditional radius interval between the two adopted \(M_{\max}\) values. The orange-shaded region shows where this conditional interval overlaps the 95\% NICER-informed reference contour; for \(M_{\max}=2.30\,M_{\odot}\), this overlap is essentially absent, whereas larger assumed \(M_{\max}\) values allow overlap for masses \(\gtrsim 2.30\,M_{\odot}\), with candidate radii extending up to \(\sim 12\,\mathrm{km}\). For completeness, the red vertical line at \(R=11\,\mathrm{km}\) indicates the maximum radius obtained from the intersection of the purple band with the \(M_{\max}=2.30\,M_{\odot}\) line \citep{IariaR_2025arXiv251120499I,BrandesL_2025PhRvD.111c4005B}. The bold cyan curve marks the radius locus obtained from Eq.~\ref{eq:MofRnum} when imposing \(M_{\max}=2.59\,M_{\odot}\) as in \citet{RochaL_2023Univ...10....3R}. The solid light-green curve, together with its dashed light-green vertical boundaries, shows the masses and radii obtained by identifying the upper kHz QPO at \(1065\,\mathrm{Hz}\) with the ISCO frequency in the Kerr metric. The solid fuchsia curve and its dashed fuchsia vertical boundaries show the analogous construction in the Hartle--Thorne metric for \(k=1\); the dash-dotted fuchsia curve shows the case \(k=8\). The hatched fuchsia region encloses the family of Hartle--Thorne \(M\)--\(R\) loci for \(k\in[1,8]\). The red-shaded region marks the \(M\)--\(R\) sector where the redshift-based locus, the EOS-informed benchmark, and the QPO/ISCO construction overlap under the adopted assumptions. It is a consistency sector, not a posterior credible region or a unique source solution.}
\label{fig:Locus}
\end{figure}

Figure \ref{fig:Locus} summarizes the comparison: we show in this figure the marginal 95\% posterior credible bands and medians for the mass-radius relation from \cite{BrandesL_2025PhRvD.111c4005B}, together  with:

\begin{itemize}
    \item In purple+orange: The region generated by the curves in Eq. \ref{eq:MofRnum} for the positive and negative uncertainties, which take into account the uncertainty in redshift, including the rotation of the NS at 716 Hz. In orange: the EOS-informed 95\% NICER-derived reference contour from \citet{BrandesL_2025PhRvD.111c4005B}. 
    \item In dark blue: The best-fit inverse compactness \(R/M = 4.46 \pm 0.13\,km/M_{\odot}\), inferred from a redshift of \(1 + z = 1.72 \pm 0.05\);
    \item Black dotted horizontal lines: at \(M_{max}=2.30\,M_{\odot}\) is the limit assumed by  \cite{IariaR_2025arXiv251120499I}, and at \(M_{max}=2.59\,M_{\odot}\) is the limit obtained by \cite{RochaL_2023Univ...10....3R};
    \item Red dotted vertical lines: the radii obtained from Equation \ref{eq:MofRnum} when imposing \(M_{\max}=2.30\,M_{\odot}\) as in \cite{IariaR_2025arXiv251120499I}. The leftmost red dotted line corresponds to the positive redshift-uncertainty branch, and the rightmost one to the negative branch.
    \item Bold cyan vertical line: the radius obtained from Equation \ref{eq:MofRnum} when imposing \(M_{\max}=2.59\,M_{\odot}\) as in \cite{RochaL_2023Univ...10....3R}.
    \item The solid green line and the two fuchsia lines (the solid and the dotted-dashed) represent the M-R relation from the QPO constraint for the Kerr and Hartle \& Thorne metrics, respectively, as discussed in section \S 2.2.
    \item The dashed vertical green and fuchsia lines delimit the mass and radii intervals.
    \item The red-shaded area marks the conditional \(M\)--\(R\) overlap sector obtained when the QPO/ISCO-based construction is compared with the redshift-implied locus and the external EoS-informed reference contour.
\end{itemize}

The contour adopted from \cite{BrandesL_2025PhRvD.111c4005B} is not a source-specific mass-radius measurement of 4U 1820-30. We use it as an EOS-informed benchmark derived from NICER-constrained neutron-star matter, i.e., as a broad consistency check rather than as an independent observational determination for this source.


The resulting purple/orange-shaded area represents the conditional \(M\)–\(R\) sector generated by Eq.~\ref{eq:MofRnum} under the adopted redshift and rotation assumptions, over the explored range \(M_{\max}=2.30\)--\(2.60\,M_\odot\). The orange portion indicates where this conditional sector overlaps the external EoS-informed NICER reference contour. For \(M_{\max}=2.30\,M_\odot\), this overlap is essentially absent or marginal when using the redshifted-Fe-line interpretation reported by Iaria et al. as input. For larger assumed values of \(M_{\max}\), the redshift-implied locus extends into the reference-contour region, with radii reaching \(\sim 12\) km. This should be read as a conditional overlap test, not as a statistically defined allowed region for 4U~1820--30.


The red area therefore marks the conditional overlap sector in the \(M\)–\(R\) plane obtained when the QPO/ISCO-based construction is overlaid on the redshift-implied locus and the external EoS-informed reference contour. Accordingly, the curves based on the QPO/ISCO approach should be read as a conditional comparison layer, not as a strict filter defining a preferred source-specific \(M\)–\(R\) region.

The red-shaded overlap region in Figure \ref{fig:Locus} is not a posterior credible region and does not define a unique solution for the source. It has a diagnostic role: it marks where the redshift interpretation, EOS-informed benchmarks, and the timing-based construction can be made mutually compatible in the \(M\)--\(R\) plane. Conversely, the small size or absence of such an overlap under a low-\(M_{\max}\) assumption is itself informative, because it exposes the tension among the redshift interpretation, EOS-informed benchmarks, and timing-based mass estimates.

We emphasize that the overlap region in Figure \ref{fig:Locus}, under the adopted assumptions, refers to candidate properties of this particular neutron star in 4U 1820-30. It should not be read as a measurement of the universal EOS maximum mass \(M_{\max}\).

Therefore, allowing a higher assumed EOS-level maximum mass, here \(M_{\max}=2.59\,M_\odot\), produces a conditional \(M\)--\(R\) overlap between the redshift-implied locus and the external EoS-informed NICER reference contour, once the adopted redshift, spin, and inclination assumptions are imposed.

The red-shaded region highlights the candidate \(M\)--\(R\) values for 4U 1820--30 obtained when the population-level interval \(M_{\max}\in[2.43,2.59]\,M_\odot\), inferred by \citet{RochaL_2021arXiv210708822R,RochaL_2023Univ...10....3R} from the observed neutron-star mass distribution, is used as an EOS-level admissibility benchmark. In this paragraph, \(M_{\max}\) is an EOS/population-level quantity: it denotes the maximum mass supported by neutron-star matter, not the mass assigned to the neutron star in 4U 1820--30.

\textbf{Note:} Figure \ref{fig:Locus} propagates only the uncertainty associated with the inferred redshift and with our rotational treatment. It does not incorporate the dominant systematics of the problem, such as line identification, formation site, continuum/absorption degeneracies, broadening assumptions, spin uncertainty, or viewing geometry. Therefore, the shaded overlap should not be interpreted as a statistically complete confidence region.

\section{Discussion}\label{discussion}

The recently proposed redshift constraint on the mass of the neutron star in 4U1820-30 has been shown to arise from a Fe line 
by \citep{IariaR_2025arXiv251120499I}. The inference was derived from a single broad, transient feature, something which necessarily relies on a set of auxiliary assumptions: the atomic identification and formation site of the feature (surface/atmosphere versus disk/wind), the adequacy of the adopted absorption model in the presence of continuum degeneracies, and the conditional interpretation of the line width in terms of rapid rotation and viewing geometry. In particular, the ``rotation wedge'' construction assumes a tentative spin frequency of 716\,Hz at $\sim$2.9$\sigma$ significance, and a specific inclination, together with a treatment of rotation that neglects higher-order effects (e.g., the quadrupole moment) and adopts approximate prescriptions for the angular momentum/moment of inertia \citep{IariaR_2025arXiv251120499I}.

A key ingredient for the former is the implicit assumption on the maximum neutron star mass $M_{\max}$: the low-$M_{\max}$ assumption adopted by \citet{IariaR_2025arXiv251120499I}, namely, $M_{\max}=2.3\,M_\odot$ following \citealt{FanY_2024PhRvD.109d3052F}, restricts the conditional sector explored in the \(M\)--\(R\) plane and drives the redshift-implied compactness into tension with the external EoS-informed NICER reference contour \citep{BrandesL_2025PhRvD.111c4005B}, as shown in Section \ref{section:Results}.

On the other hand, previous burst-based analyses of 4U 1820-30 do not define a unique source-specific mass-radius region. Rather, they span different regions because they rely on different assumptions regarding the photospheric composition, source distance, touchdown identification, spectral corrections, anisotropy, and other burst-model ingredients. The disagreement is therefore not just a matter of numerical mass estimates; it traces back to different auxiliary assumptions built into each method.

We performed a compact consistency test in the $M$--$R$ plane and reproduced the locus implied by the redshift from \cite{IariaR_2025arXiv251120499I} together with its rotation wedge construction, and overlaid it onto an external EoS-informed NICER-derived \(M\)--\(R\) reference contour from other neutron star systems. We then repeated the comparison for different assumptions about $M_{\max}$, including a high-$M_{\max}\sim \, 2.6\,M_\odot$ benchmark motivated by population-based
inferences on the mass distribution of neutron stars and by recent discussions of massive spiders \citep{RochaL_2021arXiv210708822R,RochaL_2023Univ...10....3R,BenvenutoO_2025AN....34670034B,HorvathJ_2022atcc.book....1H,HorvathJ_2023A&A...672L..11H,AlsingJ_2018MNRAS.478.1377A}. The recent report of a refined analysis of the mass of PSR J0952-0607 ($M = 2.35 \pm 0.11 M_{\odot}$, \cite{Romani}) suggests that a higher value of $M_{max}$ may be preferred, in line with the statistical inference obtained by \citet{RochaL_2023Univ...10....3R}.

We did not attempt to analyze the spectral feature itself; instead, we wanted to clarify when (and under which conditions) its redshift interpretation, advocated by the authors in \cite{IariaR_2025arXiv251120499I}, can be promoted to a strong equation-of-state constraint, and point out its divergence from former works based on different techniques.

\section{Conclusions}\label{conclusions}

As we have discussed in this work, attempts to determine the mass of the neutron star in the LMXB 4U~1820--30 span a broad range and remain method-dependent. While the analysis of \cite{OzelF_2016ApJ...820...28O} yielded $1.77 {^{+0.25}_{-0.28}} ~M_{\odot}$, several known systematics in burst-based methods may bias inferred radii and, consequently, the inferred mass-radius region, motivating continued caution in adopting any single value as definitive.

Meanwhile, the high-frequency QPOs of this source have been shown to {\it saturate} above a certain 
count rate/energy flux (e.g., \cite{Kaaret}), a feature interpreted as evidence that a maximum value of the frequency has been reached by the emitting region. This behavior has been interpreted as being consistent with the presence of the ISCO for the orbiting gas. Under that QPO/ISCO identification, the commonly used ISCO scaling yields a conditional lower limit of \(\gtrsim 2M_\odot\), with higher values obtained when the Kerr and Hartle–Thorne prescriptions adopted in this work are applied. This conditional lower limit remains in tension with the mass determination obtained from the corrected touchdown-flux analysis.

This situation is not improved by the latest derivation from the $\sim$ 3.8~keV feature as a gravitationally redshifted iron line formed at (or very near) the stellar surface, since the high compactness, equivalently the low inverse compactness \(R/M\simeq4.46\pm0.13\,\mathrm{km}\) \citep{IariaR_2025arXiv251120499I}, automatically points toward a large mass for radii in the usual neutron-star range. Our comparison showed that, when a low maximum mass is imposed a priori (e.g. \(M_{\max}=2.3\,M_{\odot}\), as adopted in \citealt{IariaR_2025arXiv251120499I}), the redshift-implied locus exhibits only marginal overlap with the external EoS-informed NICER reference contour derived from several neutron-star systems. Allowing a higher \(M_{\max}\) (e.g. up to \(\sim 2.6\,M_{\odot}\)) opens up parameter space in which the redshift-implied locus can intersect that reference contour, reducing the apparent inconsistency, but it makes the tension with the other determinations even worse.

A central point of this work is that the three methods do not simply yield different numerical values for the mass of the neutron star in the 4U 1820-30 system; they become mutually difficult to reconcile once their assumptions are placed in the same \(M\)--\(R\) plane. The recent report by \citet{Romani} that neutron-star masses as high as \(2.46\,M_\odot\) may be allowed within the error bars further motivates considering a high-\(M_{\max}\) sector, although it does not by itself resolve the source-specific tension in 4U 1820--30.

We emphasize that this work does not determine the mass of 4U 1820--30, nor does it infer \(M_{\max}\) from this source. The source mass \(M\) and the EoS/population-level maximum mass \(M_{\max}\) play different roles in our argument: \(M\) labels candidate stellar masses in the \(M\)--\(R\) plane, whereas \(M_{\max}\) is used only as an external admissibility scale. This distinction is relevant because the theoretical upper bound on \(M_{\max}\) is itself assumption-dependent. The classical Rhoades--Ruffini value of \(3.2\,M_{\odot}\) relies on assumptions about the high-density EoS \citep{RhoadesRuffini1974}, and recent work has shown that, if a stiff high-density phase sets in near or below nuclear saturation density, the maximum mass of hybrid neutron stars may reach \(4\,M_{\odot}\) or higher \citep{BlaschkeWojcik2026}. Thus, the use of \(M_{\max}\) in this paper should not be read as assigning the mass of 4U 1820--30 to the maximum mass allowed by the EoS, nor as imposing a low fixed ceiling on neutron-star masses.

The main result of this work is the identification of a specific tension in the \(M\)--\(R\) plane. If the 3.8 keV feature is interpreted as a surface or near-surface redshifted Fe line, the implied compactness is hard to reconcile with a low maximum neutron-star mass, such as \(M_{\max}=2.3\,M_\odot\). A population-level value closer to \(2.6\,M_\odot\) restores overlap with the EOS-informed NICER benchmark, but it also sharpens the disagreement with burst-based mass estimates. Thus, the comparison does not simply add another mass estimate for 4U 1820--30; it shows which assumptions drive the apparent compatibility, or incompatibility, among the redshift, burst, and QPO/ISCO interpretations.

We have shown that using the reported transient feature as an equation-of-state constraint is sensitive to multiple high-impact auxiliary assumptions, e.g., line identification/formation site, continuum/absorption modeling degeneracies, and the adoption of rapid rotation and viewing geometry. Additional observations confirming the feature and improved modeling are needed to resolve these issues and sharpen the inferred neutron star parameters in this system. Even without considering the new line, the former determinations are in conflict with one another, an issue that, if properly clarified, would be very helpful for the determination of the neutron star masses in this class of binaries by itself.

This comparison also makes clear what a future joint inference would have to include: the line identification and formation site, the burst-model systematics, the QPO/ISCO assumption, the spin, the inclination, and the EOS-level prior on \(M_{\max}\).

\section{Acknowledgements}
L.S.R acknowledges the São Paulo Research Foundation (FAPESP) through the grants \#2025/05690-2 and \#2023/08649-8. J.E.H. acknowledges the continued support of the CNPq Federal Agency, Brazil, over the years. The São Paulo Research Foundation (FAPESP) supported the activities of the authors through the grant \#2024/16892-2. L.M.S. acknowledges support by the Alexander von Humboldt Foundation. M.G.B.A. acknowledges the São Paulo Research Foundation (FAPESP) through the grant \#2025/25588-8. The authors would like to thank the Brazilian Ministry of Science, Technology and Innovation (MCTI) and the Brazilian Space Agency (AEB), which supported the present work under the PO 20VB.0009.

\section{Declaration of generative AI and AI-assisted technologies in the manuscript preparation process.}

During the preparation of this work, the author(s) used ChatGPT Plus to keep the Abstract under the limit of 250 words. After using this tool/service, the author(s) reviewed and edited the content as needed and take(s) full responsibility for the content of the published article.

The author(s) used Google NotebookLM to create an image for the graphical abstract. After using this tool/service, the author(s) reviewed and edited the content as needed and take(s) full responsibility for the content of the published article.

\printcredits

\bibliographystyle{cas-model2-names}

\bibliography{cas-refs}






\end{document}